# Stochastic perturbations to dynamical systems: a response theory approach

Valerio Lucarini

Klimacampus, Institute of Meteorology, University of Hamburg, Hamburg, Germany

[valerio.lucarini@zmaw.de]

Department of Mathematics and Statistics, University of Reading, Reading, UK

## Abstract

We study the impact of stochastic perturbations to deterministic dynamical systems using the formalism of the Ruelle response theory and explore how stochastic noise can be used to explore the properties of the underlying deterministic dynamics of a given system. We find the expression for the change in the expectation value of a general observable when a white noise forcing is introduced in the system, both in the case of additive and multiplicative noise. We also show that the difference between the expectation value of the power spectrum of an observable in the stochastically perturbed case and of the same observable in the unperturbed case is equal to the variance of the noise times the square of the modulus of the susceptibility describing the frequency-dependent response of the system to perturbations with the same spatial patterns as the considered stochastic forcing. Using Kramers-Kronig theory, it is then possible to derive the real and imaginary part of the susceptibility and thus deduce the Green function of the system for any desired observable. We then extend our results to rather general patterns of random forcing, from the case of several white noise forcings, to noise terms with memory, up to the case of a space-time random field. Explicit formulas are provided for each relevant case analysed. As a general result, we find, using an argument of positive-definiteness, that the power spectrum of the stochastically perturbed system is larger at all frequencies than the power spectrum of the unperturbed system. We provide a example of application of our results by considering the spatially extended chaotic Lorenz 96 model. These results clarify the property of stochastic stability of SRB measures in Axiom A flows, provide tools for analysing stochastic parameterisations and related closure ansatz to be implemented in modelling studies, and introduce new ways to study the response of a system to external perturbations.

# 1. Introduction

In many scientific fields, numerical modelling is taking more and more advantage of supplementing traditional deterministic numerical models with additional stochastic forcings. This has served the overall goal of achieving an approximate but convincing representation of the spatial and temporal scales which cannot be directly resolved. Moreover, stochastic noise is usually thought as a reliable tool for accelerating the exploration of the attractor of the deterministic system, by the additional mixing due to noise, and, in some heuristic sense, to increase the robustness of the model, by getting rid of potentially pathological solutions. This may be especially desirable when computational limitations hinder our ability to perform long simulations. For both of these reasons, climate science is probably the field where the application of the so-called stochastic parameterisations is presently gaining more rapidly momentum for models of various degrees of complexity, including full-blown GCMs; see, e.g. [1], after having enjoyed an early popularity for enriching and increasing the realism of very simple models with few degrees of freedom [2].

Previous works have addressed the problem of computing the linear response of a nonlinear system to perturbation in presence of a background stochastic forcings. See [3-5] for notable examples of theoretical findings and algorithms. Nonetheless, these contributions basically focus on developing response formulas for short and long term horizon aimed at describing the impact of changing the deterministic part of the evolution operator of the dynamical system, while the stochastic component is a given forcing which contributes to determining the invariant measure. Note that adding stochastic noise partly simplifies the mathematics related to the formal derivation of the response because, as opposed to what is usually the case for chaotic deterministic systems, the invariant measure is absolutely continuous with respect to Lebsesgue, so that, e.g., the fluctuation-dissipation theorem applies [6].

Instead, in this paper we take a rather different and complementary point of view. We wish to provide some new analytical results with a rather large degree of generality on the impact of adding stochastic forcings to deterministic system, with the goal of possibly providing useful guidance for closure problems related to large scale application of stochastic parameterisation. We will study this problem by taking advantage of the response theory developed by Ruelle [7-9] for studying the impact of small perturbations to rather general flows. Whereas the theory has been developed for deterministic perturbations, we will adapt its results to perturbations which have a stochastic nature and derive analytical expressions describing how noise changes the expectation value of a general observable. In other terms, this amounts to deriving how the invariant measure changes as a result of adding the stochastic perturbation. We will also derive new results which show how the changes in the spectral properties of the system due to introduction of the stochastic perturbations can be used to derive the general properties on the frequency-dependent response of the system – more precisely its susceptibility [9-11] - thus providing information on the fine structure of the attractor. This is accomplished by applying Kramers-Kronig theory [10,11] to suitably defined spectral functions. The paper is organised as follows. In Section 2 we present the theoretical background of our investigation, showing how response theory can be used to study the impact of stochastic perturbations. We also present some explicit results for the simplest case of random forcing considered in this study, where a single white noise perturbation is considered. In Section 3 we extend our analysis to more and more general patterns of random forcing, up to the case of a general space-time random field. We present explicit formulas for various relevant intermediate cases. In Section 4 we present the result of a numerical experiment conducted with the classical Lorenz 96 model of the atmosphere [12,13], in order to demonstrate the great potential and the applicability of the results presented here. In section 5 we present our conclusions and perspective for future work.

## 2. Response Theory and Stochastic Perturbations

Let's frame our problem in a mathematically convenient framework. Axiom A dynamical systems of the form $dx_i/dt = F_i(x)$ (all of our results can be easily reframed for discrete maps) possess a very special kind of invariant measure $\rho_0(dx)$, usually referred to as SRB measure. Such a measure is first of all a physical measure, i.e. for a set of initial conditions of positive Lebesgue measure the time average $\lim_{T \to \infty} 1/T \int_0^T dt A(f^t x)$ of any smooth observables $A$, with $f$ being the evolution operator of the flow, converges to the expectation value $\rho_0(A) = \int \rho_0(dx) A(x)$. Another remarkable property of $\rho_0(dx)$ is that it is stochastically stable, i.e. it corresponds to the zero noise limit of the invariant measure of the random dynamical system whose zero-noise is the deterministic system $dx_i/dt = F_i(x)$. See [14-16] for a much precise and detailed description of Axiom A systems and SRB measures.

Ruelle [7-9] (see also [17] for a slightly different point of view) has shown that one can construct a response theory able to express the change in the expectation value $\delta_{\varepsilon,t}\rho(A) = \rho_{\varepsilon,t}(A) - \rho_0(A)$ of an arbitrary measurable observable A when the flow undergoes a small perturbation of the form $dx_i/dt = F_i(x) \to dx_i/dt = F_i(x) + \varepsilon X_i(x)g(t)$, where $X_i(x)$ is a smooth vector field, $g(t)$ is its time modulation and $\varepsilon$ is the order parameter of the perturbation (we introduce such a factor in order to clarify the perturbative expansion). The main result is that $\delta_{\varepsilon,t}\rho(A)$ can be written as a Kubo-type power series [18]:

$$\delta_{\varepsilon,t}\rho(A) = \varepsilon \int d\tau G_1(\tau) g(t-\tau) +$$
$$+ \varepsilon^2 \int d\tau_1 d\tau_2 G_2(\tau_1,\tau_2) g(t-\tau_1) g(t-\tau_2) + \sum_{k=3}^{\infty} \varepsilon^k \int d\tau_1 ... d\tau_k G_k(\tau_1,...,\tau_k) g(t-\tau_1) ... g(t-\tau_k) =$$
$$= \varepsilon \int \rho_0(dx) \int d\tau \Theta(\tau) X_i \partial_i A(f^\tau x) g(t-\tau) +$$
$$+ \varepsilon^2 \int \rho_0(dx) \int d\tau_1 d\tau_2 \Theta(\tau_1) \Theta(\tau_2 - \tau_1) X_i \partial_i \left( f^{\tau_2 - \tau_1} \left( X_j \partial_j A(f^{\tau_1} x) \right) \right) g(t-\tau_1) g(t-\tau_2) + o(\varepsilon^3)$$

(1)

where $\partial_i(*) = \partial(*)/\partial x_i$ and $f^\tau$ is the evolution operator of the unperturbed flow. $G_k(\tau_1,...,\tau_k)$ is the $k^{th}$ order Green function of the system, which, convoluted $k$ times with the delayed modulation function $g(t)$, describes the contribution to the response resulting from interaction at the $k^{th}$ order of nonlinearity of the perturbation. In Eq. (1), we present the first two terms of the series $G_1(\tau_1) = \int \rho_0(dx) \Theta(\tau_1) X_i \partial_i A(f^{\tau_1} x)$ and $G_2(\tau_1,\tau_2) = \int \rho_0(dx) \Theta(\tau_1) \Theta(\tau_2 - \tau_1) X_i \partial_i \left( f^{\tau_2 - \tau_1} \left( X_j \partial_j A(f^{\tau_1} x) \right) \right)$. Two fundamental properties of formula given in Eq. (1) are that at all orders 1) the Green function is causal, and 2) the contribution to the response is written as expectation value over the unperturbed state of a an observable for which we have an explicit expression [7-9, 18]. Since $\rho_0(dx)$ is a physical measure, in practical terms the Green function can be evaluated by producing long numerical integration of the dynamical system $dx_i/dt = F_i(x)$ and evaluating suitably the time averages. In this direction, efficient algorithms have been proposed, e.g. in [17]. Unfortunately, the models describing the dynamics of natural or artificial systems are typically not Axiom A, so that the Ruelle response theory may appear to be of little practical interest. Nonetheless, the chaotic hypothesis [19] suggests that, when we consider systems with many degrees of freedom and analyse smooth observables, the behaviour is typically close to that of Axiom A systems. Therefore, a much wider applicability beyond the mathematical limits reached so far is reasonable. As an example, numerical evidences suggest that Ruelle's theory and its extension in the frequency domain [18] provides consistent results for the linear and nonlinear response of the Lorenz 63 system [20] and the Lorenz 96 system [21]. Furthermore, various additional results obtained by Majda and collaborators suggest a vast range of applicability of response theory and a clear stability of the algorithms used to study it [17,22,23].

**2.a Basic result: single white noise perturbation**

We now assume that the perturbation is modulated by white noise, so that $g(t) = \eta(t) = dW(t)/dt$, where $W(t)$ is a Wiener process. Therefore, $\langle \eta(t) \rangle = 0$ and $\langle \eta(t)\eta(t') \rangle = \delta(t-t')$, where the symbol $\langle \bullet \rangle$ describes the operation of averaging over the realizations of the stochastic processes. Note that, since the invariant measure $\rho_0(dx)$ is stochastically stable, as discussed above, it makes sense to compute the impact of weak stochastic perturbations. Since we are actually dealing with a stochastic dynamical system, we redefine our response as $\langle \delta_{\varepsilon,t}\rho(A) \rangle$. When the statistical properties of $\eta(t)$ are taken into consideration, we obtain the following formula for the response of the system by considering the expectation value of Eq. (1):

$$\langle \delta_\varepsilon \rho(A) \rangle = \varepsilon^2 \int d\tau_1 G_2(\tau_1,\tau_1) + o(\varepsilon^4) = 1/2\,\varepsilon^2 \int \rho_0(dx) \int d\tau_1 \Theta(\tau_1) X_i \partial_i X_j \partial_j A(f^{\tau_1}x) + o(\varepsilon^4). \quad (2)$$

All odd order terms are vanishing because of the symmetry properties of the noise, and the leading term results to be proportional to the square of the order parameter $\varepsilon$ times the time-independent expectation value of an observable on the unperturbed measure. The factor ½ emerges from evaluating $\Theta(\tau)$ taking a symmetric limit for $\tau \to 0$. As expected, the response has only a static component, as no time-dependence is present. Moreover, the fact that $\langle \delta_\varepsilon \rho(A) \rangle$ depends smoothly on the intensity of noise and vanishes for noise of vanishing intensity is in agreement with the stochastic stability of the SRB measure $\rho_0(dx)$.

We take a brief diversion to the special case of non-singular invariant measures. If the unperturbed invariant measure $\rho_0(dx)$ is absolutely continuous with respect to Lebesgue $\rho_0(dx) = \rho_0(x)dx$, as in the case of equilibrium systems, by performing an integration by parts described in [7-9], we obtain for the leading term:

$$\langle \delta_\varepsilon \rho(A) \rangle \approx -\varepsilon^2 \int d\tau \Theta(\tau_1) \int dx \rho_0(x)(\partial_i \bullet X_i) X_j \partial_j A(f^{\tau_1}x) \quad (4)$$

where $\partial_i \bullet X_i = 1/\rho_0 \, \partial_i(\rho_0 X_i)$ [7], so that the leading term can be written as proportional to the integral in the positive time domain of the correlation of two suitably defined observables *B* and *C*:

$$\langle \delta_\varepsilon \rho(A) \rangle \approx -\varepsilon^2 \int d\tau \Theta(\tau) \int \rho_0(dx) B(x) C(f^\tau x) \tag{5}$$

where $B = \partial_i \bullet X_i = 1/\rho_0 \, \partial_i(\rho_0 X_i)$, while $C = X_j \partial_j A(f^{\tau_1} x)$ is such that its expectation value is exactly the linear the Green function of the system - see Eq. (1). This can be interpreted as a variant of the fluctuation-dissipation theorem.

Let's now consider the ensemble average over the probability space of the considered stochastic processes of the expectation value of the time correlation of the response of the system. We then consider the following expression:

$$\begin{aligned}\left\langle \int d\sigma \delta_{\varepsilon,\sigma}\rho(A) \delta_{\varepsilon,\sigma-t}\rho(A) \right\rangle &= \varepsilon^2 \int d\sigma d\tau_1 d\tau_2 G_1(\sigma - \tau_1) G_1(\sigma - t - \tau_2) \langle \eta(\tau_1)\eta(\tau_2) \rangle + o(\varepsilon^4) \\ &= \varepsilon^2 \int d\sigma G_1(\sigma) G_1(\sigma - t) + o(\varepsilon^4) \end{aligned} \tag{6}$$

where, as shown in Eq. (1), $G_1(\tau) = \int \rho_0(dx) \int d\tau \Theta(\tau) X_i \partial_i A(f^\tau x)$. By applying the Fourier transform to both members and using the Wiener-Khinchin theorem, we obtain:

$$\left\langle \left| \delta_{\varepsilon,\omega}\rho(A) \right|^2 \right\rangle \approx \varepsilon^2 |\chi_1(\omega)|^2, \tag{7}$$

where the susceptibility $\chi_1(\omega) = \Phi_\omega[G_1(t)]$ is the Fourier Transform of the $G_1(t)$ and we have neglected the higher-order terms. Therefore, the ensemble average of the power spectrum $\left| \delta_{\varepsilon,\omega}\rho(A) \right|^2 = \left| \Phi_\omega[\delta_{\varepsilon,t}\rho(A)] \right|^2$ of the response to the random perturbation is, just like in the case of equilibrium systems (see e.g. the explicit expression for the forced and damped linear oscillator [11]) approximately proportional to the square of the modulus of the linear susceptibility via the square of the order parameter $\varepsilon$. Considering the ergodic nature of the unperturbed flow and the fact that the stochastic perturbation is a white noise, after some algebraic manipulations we derive that:

$$\langle P_{\varepsilon,\omega}(A)\rangle - P_{\omega}(A) \approx \langle |\delta_{\varepsilon,\omega}\rho(A)|^2\rangle \approx \varepsilon^2 |\chi_1(\omega)|^2 \qquad (8)$$

where $P_{\omega}(A) = \Phi_{\omega}\left[\int \rho_0(dx)A(x)A(f^t x)\right]$ is the power spectrum of the observable A in the unperturbed flow while $P_{\varepsilon,\omega}(A) = \Phi_{\omega}\left[\lim_{T\to\infty}(1/T)\int_0^T dt A(x)A(f_\varepsilon^t x)\right] \approx \Phi_{\omega}\left[\int \rho_0(dx)A(x)A(f_\varepsilon^t x)\right]$ is the power spectrum of the perturbed flow – with $f_\varepsilon^t$ indicating the corresponding evolution operator; the approximate identity is valid in the weak noise limit. Therefore, up to the second order in $\varepsilon$, the difference between the expectation value of the power spectrum of the observable A of the perturbed flow and the power spectrum of the observable A in the unperturbed flow gives the ensemble average of the power spectrum of the response, which is proportional to the square of the modulus of the susceptibility. Equation (8) is much more useful than Eq. (7) because the left hand side member can be observed much more easily. By measuring experimentally the difference between the power spectrum of an observable in the presence of noise (for several realizations of the noise) and in absence of noise, it is possible to reconstruct the square of modulus of the susceptibility of the system.

As widely discussed in [7-9,18,20,21], the susceptibility is an analytic function in the upper complex $\omega$-plane, so that it obeys Kramers-Kronig relations [10,11]. Using analyticity, we can derive $\chi_1(\omega)$ from $|\chi_1(\omega)|$. In fact, we can write $\chi_1(\omega) = |\chi_1(\omega)|e^{i\varphi(\omega)}$ and, by taking the logarithm to both members, we obtain $\log[\chi_1(\omega)] = \log[|\chi_1(\omega)|] + i\varphi(\omega)$. The function $\log[\chi_1(\omega)]$ is also analytic in the upper complex $\omega$-plane, so that it also obeys Kramers-Kronig relations. Therefore, from the knowledge of the real part for all values of $\omega$ we can obtain the value of its imaginary par via a Hilbert transform (and vice versa). In this case, the real part is given by $\log[|\chi_1(\omega)|] = 1/2\log[|\chi_1(\omega)|^2]$, which can be derived by the analysis of the power spectra in the perturbed and unperturbed case using Eq. (8), whereas the imaginary part, obtained using Kramers-Kronig relations, will give the phase function $\varphi(\omega)$. Note that such standard reconstruction technique is widely used in optics to derive the index of refraction of a material from its reflectivity [10]. The strategy of taking the logarithm of the susceptibility falls into some troubles if the function $G_1(\omega)$ possesses some zeros in the upper complex $\omega$-plane. At any rate, both exact and approximate numerical techniques have been devised to take care of this rather special case [10,11]. Once we have obtained $\varphi(\omega)$, we can reconstruct $\chi_1(\omega)$, and thus derive full information on the linear response of the system to perturbations with the same spatial pattern and with general temporal modulation. The Green function of the system $G_1(t)$ can be obtained using an inverse Fourier transform and, as discussed in [21], it can be used to predict finite and infinite time horizon response of the system to perturbations with the same spatial structure but general time modulation.

## 3 Extension to more general cases of random perturbations

### 3.a Case of several white-noise perturbations

We now analyse the case of a more complex pattern of stochastic forcing, so that the perturbation can be written as $dx_i/dt = F_i(x) \to dx_i/dt = F_i(x) + \sum_{j=1}^{p} \varepsilon_j X_i^j(x) \eta_j(t)$, where we consider $p$ independent perturbative vector flows $X^j(x)$ and $p$ white noise signals $\eta_j(t)$ such that $\langle \eta_j(t) \rangle = 0$ $\forall j$ and $\langle \eta_i(t) \eta_j(t') \rangle = C_{ij} \delta(t-t')$, where $C_{ij} = C_{ji}$ is the cross-correlation matrix which has unitary entries in the diagonal. Under these conditions, it is straightforward to prove that:

$$\langle \delta_{\{\varepsilon\}} \rho(A) \rangle = 1/2 \sum_{l,m=1}^{p} \varepsilon_l \varepsilon_m C_{lm} \int \rho_0(dx) \int d\tau_1 \Theta(\tau_1) X_i^l \partial_i X_j^m \partial_j A(f^{\tau_1} x) + o(\{\varepsilon\}^4) \tag{9}$$

By exploiting the symmetry of the correlation matrix, Eq. (9) can be rewritten as follows:

$$\langle \delta_{\{\varepsilon\}} \rho(A) \rangle = \sum_{k=1}^{p} \varepsilon_k^2 \int d\tau_1 G_2^{k,k}(\tau_1,\tau_1) + \sum_{l>m=1}^{p} \varepsilon_l \varepsilon_m C_{lm} \int d\tau_1 G_2^{l,m}(\tau_1,\tau_1) + o(\{\varepsilon\}^4) \tag{10}$$

where the second order Green functions corresponding to the joint effect of the $l^{th}$ and $m^{th}$ perturbation vector fields are written as:

$$G_2^{l,m}(\tau_1,\tau_1) = 1/4 \int \rho_0(dx) \int d\tau_1 \Theta(\tau_1)(2-\delta_{l,m})\left(X_i^l \partial_i X_j^m + X_i^m \partial_i X_j^l\right) \partial_j A(f^{\tau_1} x) \tag{11}$$

If $C_{ij} = \delta_{ij}$, so that the $p$ white noise processes are uncorrelated, the second term on the right hand side of Eq. (10) vanishes. In this case, the contributions of the various stochastic perturbations sum up linearly (weighted by their variance) even if we are considering second order effects.

When considering the power spectrum of the response, we obtain the following generalisation of Eq. (8):

$$\langle P_{\{\varepsilon\},\omega}(A)\rangle - P_\omega(A) \approx \langle |\delta_{\{\varepsilon\},\omega}\rho(A)|^2 \rangle \approx \sum_{l=1}^{p} \varepsilon_k^2 |\chi_1^k(\omega)|^2 + \\ + \sum_{l>m=1}^{p} \varepsilon_l \varepsilon_m C_{l,m} \left[\chi_1^l(\omega)(\chi_1^m(\omega))^* + (\chi_1^l(\omega))^* \chi_1^m(\omega)\right] \quad (12)$$

where $\chi_1^l(\omega)$ is the Fourier Transform of $G_1^l(\tau) = \int \rho_0(dx)\int d\tau \Theta(\tau) X_i^l \partial_i A(f^\tau x)$. Since by definition the matrix $C_{ij}$ is positive semi-definite, the right hand side of Eq. (12) under no circumstances can become negative, thus generalising the result on the positive sign of the difference between the perturbed and unperturbed power spectra discussed in Eq. (8). If $C_{ij} = \delta_{ij}$, the impact of the various stochastic forcings sum up linearly, so that the square moduli of the various Green functions are weighted with the square of the intensity of the noise. If we have the possibility of varying the intensity of the various noise components, we can disentangle the square modulus of each Green functions and, as described above, eventually of the full Green function.

Note that, as opposed to what discussed in the previous Section, the application of the Kramers-Kronig relations to the logarithm of the right hand side member of Eq. (12) cannot in general be used to derive the real and imaginary part of the "effective" linear susceptibility $\sum_{k=1}^{p} \varepsilon_k \chi_1^k(\omega)$ corresponding to a perturbation of the vector field expressed as $F_i(x) \to F_i(x) + g(t)\sum_{j=1}^{p} \varepsilon_j X_i^j(x)$. This is so unless, e.g. all susceptibilities are identical as in the case of presence of special symmetry properties in the forcings and in the system.

**3.b Case of a single general random perturbation**

We now assume that the perturbation is modulated by a single, but rather general random perturbation, so that $dx_i/dt = F_i(x) \to dx_i/dt = F_i(x) + X_i(x)\eta(t)$ where $\langle \eta(t) \rangle = 0$ and $\langle \eta(\tau_1)\eta(\tau_2) \rangle = D(\tau_2 - \tau_1)$, with $\int_{-\infty}^{\infty} d\tau D(\tau) = \varepsilon^2$. Plugging these properties into Eq. (1), we immediately derive the following result at leading order in the intensity of the noise:

$$\langle \delta_\varepsilon \rho(A) \rangle \approx \int \rho_0(dx) \int d\tau_1 d\sigma \Theta(\tau_1) \Theta(\sigma) X_i \partial_i \left( f^\sigma \left( X_j \partial_j A(f^\tau x) \right) \right) D(\sigma) \tag{13}$$

where the function $D(\sigma)$ has the role of a memory term, weighting the various delayed 2$^{nd}$ order contributions. As in the previous cases, the expectation value of a generic observable $A$ is changed by a constant we can explicitly compute. Following closely the same calculation given in subsection 2.a, we obtain:

$$\langle P_{\varepsilon,\omega}(A) \rangle - P_\omega(A) \approx \langle |\delta_{\varepsilon,\omega}\rho(A)|^2 \rangle \approx |\chi_1(\omega)|^2 D(\omega) \tag{14}$$

where $D(\omega) = \Phi_\omega[D(\tau)]$. Therefore, if the time correlation (or equivalently, the spectral) properties of the noise input are known, it is still possible to obtain the squared modulus of the susceptibility by observing at the difference between the power spectrum of the considered observable in the two scenarios with and without noise. Moreover, since $D(\tau)$ is Hermitian symmetric and positive semidefinite, it is possible to write $D(\omega)$ as $D(\omega) = H(\omega)(H(\omega))^* = |H(\omega)|^2$ under the rather mild condition that the Paley-Wiener criterion [24] applies. Therefore, for any noise signal $\eta(\tau)$, the left hand side of Eq. (12) is positive, thus implying that in all generality the power spectrum of any observable is larger in the presence than in the absence of noise for all values of $\omega$.

We note that it is possible to find an explicit expression of $H(\omega)$ under the hypothesis that $D(\omega)$ is a rational function by imposing a minimum phase condition [25]. A specific example of direct evaluation of $H(\omega)$ can be provided as follows. Let's assume that the random forcing $\eta(\tau)$ results from (or at least can be modelled as resulting from) a linear process of the form $L[\eta(\tau)] = w(\tau)$, where $L[\bullet]$ is a differential operator and $w(\tau)$ is a white noise of variance $\delta^2$. These are continuous time equivalent of AR(n) processes. In this special case we have that - $H(\omega) = \delta/L(i\omega)$, where the differential $L$ operator is interpreted as an algebraic polynomial. Note that this example is of practical relevance because some of most efficient ways of estimating the spectral properties of a signal (e.g. Yule-Walker and Maximum Entropy/Burg methods), boil down to fitting the time series with an AR(n) model [11,26].

Note that, in the limit of delta-correlated noise, Eq. (13) agrees with Eq. (2), and Eq. (14) agrees with Eq. (7).

**3.c Case of several general random perturbations**

We further extend our analysis by considering now the case where the perturbation flow can be written as $\sum_{j=1}^{p} X_i^j(x)\eta_j(t)$, where the random signals $\eta_j(t)$ are such that $\langle \eta_j(t) \rangle = 0 \ \forall j$ and $\langle \eta_i(t)\eta_j(t') \rangle = D_{ij}(t-t')$, where $D_{ij}(\tau) = D_{ji}(-\tau)$ is the cross-correlation, and $\int_{-\infty}^{\infty} d\tau R_{ij}(\tau) = \varepsilon_i \varepsilon_j$. Whereas extending Eq. (13) is cumbersome and straightforward, we report the generalisation of Eqs. (12) and (14):

$$\langle P_{\varepsilon,\omega}(A) \rangle - P_\omega(A) \approx \langle |\delta_{\{\varepsilon\},\omega}\rho(A)|^2 \rangle \approx \sum_{l=1}^{p} |\chi_1^l(\omega)|^2 D_{l,l}(\omega) + \\ + \sum_{l>m=1}^{p} D_{l,m}(\omega)\chi_1^l(\omega)[\chi_1^m(\omega)]^* + D_{m,l}(\omega)[\chi_1^l(\omega)]^* \chi_1^m(\omega) \tag{15}$$

where $D_{l,m}(\omega) = F[D_{l,m}(\tau)]$, with the property that $D_{l,m}(\omega) = [D_{m,l}(\omega)]^*$. The semipositiveness of the cross-correlation ensures, also in this case, that the right hand side of the above equation is positive in all generality. This formula provides an explicit link between the spectral properties of the noisy forcing and the change in the power spectrum of the considered observable. Joining on to the discussion presented in subsection a, we provide an example of an explicit formula for $D_{l,m}(\omega)$. If the random signals $\eta_j(\tau)$ result from (or at least can be modelled as resulting from) linear processes of the form $L_j[\eta_j(\tau)] = w_j(\tau)$, where $L_j[\bullet]$ is a differential operator, $\langle w_j(\tau) \rangle = 0$ $\forall j$ $\langle w_i(t), w_j(t') \rangle = K_{ij}\delta(t-t')$, we obtain that $D_{i,j}(\omega) = K_{ij}(L_i(i\omega))^{-1}[(L_j(i\omega))^{-1}]^*$ where the Einstein convention on summing indices is not taken.

**3.d Extension to the case of a general random field**

We now analyse the rather general case where no assumptions are made on the coupling between the spatial and temporal patterns of the additional forcing. Therefore, the perturbation can be written as $dx_i/dt = F_i(x) \rightarrow dx_i/dt = F_i(x) + X_i(x,t)$, where $X_i(x,t)$ is a stationary random field. Under mild hypothesis of regularity, it is always possible to perform a Schauder decomposition [27] such that $X_i(x,t) = \sum_k \eta_k(t)\psi_i^k(x)$, where $\psi_i^k(x)$ is the $i^{th}$ component of the the $k^{th}$ element of the complete base defined by the decomposition. Once a decomposition is chosen, the (random) time factor $\eta_k(t)$ is obtained by projecting the perturbation field $X(x,t)$ on the the $k^{th}$ element of the complete basis at all times. Repeating this operation for all the $\eta_k(t)$'s, we can construct the covariance $D_{ij}(\tau) = D_{ji}(-\tau) = \langle \eta_i(t)\eta_j(t-\tau)\rangle$. We are basically in the same situation as described in the previous subsection, the main differences being that we have an infinite sum, because the basis has an infinite number of elements, and that the spatial patterns of perturbation are determined by the specifically chosen decomposition. We thus obtain:

$$\langle P_{\varepsilon,\omega}(A)\rangle - P_\omega(A) \approx \langle |\delta_{\{\varepsilon\},\omega}\rho(A)|^2 \rangle \approx \sum_{l=1} |\chi_1^l(\omega)|^2 D_{l,l}(\omega) +$$
$$+ \sum_{l>m=1} D_{l,m}(\omega)\chi_1^l(\omega)[\chi_1^m(\omega)]^* + D_{m,l}(\omega)[\chi_1^l(\omega)]^*\chi_1^m(\omega) \quad (16)$$

where $\chi_1^l(\omega)$ is the Fourier Transform of $G_1^l(\tau) = \int \rho_0(dx)\int d\tau\Theta(\tau)\psi_i^l\partial_i A(f^\tau x)$, and, as discussed before, the properties of the covariance ensure the non-negativity of the right hand side member of Eq. (16). Note that, whereas the various terms of the sum in Eq. (16) depend on the selected Schauder basis, by completeness, the full sum will be independent of the selected decomposition. As in practice one has to deal with finite sums, selecting a practically suitable basis expansion and its efficient truncation will impact the quality of the estimate for the left hand side member in Eq. (16). This formula includes, as special cases, Eqs. (8), (12), (14), and (15).

## 4. A Numerical Experiment

In order to provide further support for our findings, we provide a simple but nontrivial example of numerical investigation along the lines detailed above, sticking to the basic cases discussed in subsection 2.a. We consider the Lorenz 96 model [12,13], which, describes the evolution of a generic atmospheric variable defined in N equally spaced grid points along a latitudinal circle and provides a simple, unrealistic but conceptually satisfying representation of some basic atmospheric processes, such as advection, dissipation, and forcing. This model has a well recognised prototypical value in data assimilation [28, 29], predictability studies [30,31] and has been investigated in detail in terms of linear response theory [3,4,21]. The model is defined by the differential equations $dx_i/dt = x_{i-1}(x_{i+1} - x_{i-2}) - x_i + F$ with i=1,...,N and the index *i* is cyclic so that $x_{i-N} = x_i = x_{i+N}$. In order to provide results directly comparable with obtained in [21], we perturb the system with an additive white noise, so that $F \to F + \varepsilon\eta(t)$ for all grid points *i*. We take as observables the "intensive energy" $e = 1/(2N)\sum_{j=1}^{N} x_i^2$ and "intensive momentum" $m = 1/N \sum_{j=1}^{N} x_i$ of the system and consider the corresponding linear susceptibilities $\chi_{e,N}^{(1)}(\omega)$ and $\chi_{m,N}^{(1)}(\omega)$. We choose standard conditions ($F = 8.0$, $N = 40$, see discussion in [21] explaining how results can be generalized for all values of *F* and *N* as long as the system is chaotic), select $\varepsilon = 0.5$ and integrate each of the 1000 members of the ensemble of realisations of the stochastic process for 1000 time units, which correspond to about 5000 days [12,13]. We choose a computationally very inexpensive experiment (all runs have been completed in less than one day on a commercial laptop using MATLAB®) and use a rather sub-optimal way to estimate the power spectra, such as taking the square of the fast Fourier transform of the signal, in order to prove the robustness of our approach. The obtained results do not depend on the intensity of the noise for, e.g., $\varepsilon \leq 1$, except that performing simulations using a larger value of $\varepsilon$ improves the signal-to-noise ratio. The squared modulus of the susceptibility $\left|\chi_{e,N}^{(1)}(\omega)\right|^2$ obtained using Eq. (8) (blue line) and its direct estimate drawn from the data for published in [21] (black line) are in excellent agreement, so that in Figure 1a we shift vertically one of the two curves to improve readability. We also plot (red line) the high-frequency asymptotic behaviour $\left|\chi_{e,N}^{(1)}(\omega)\right|^2 \approx \rho_0(m)^2/\omega^2$ derived analytically using the short-time expansion of the Green function in [21], in order to show how accurately the methodology presented in this paper is able to capture the response of the system to high frequency perturbations. Similarly, in Figure 1b) we present the corresponding results for $\left|\chi_{m,N}^{(1)}(\omega)\right|^2$, which feature a comparable degree of accuracy throughout the spectral range. In this case, the asymptotic behaviour is $\left|\chi_{m,N}^{(1)}(\omega)\right|^2 \approx 1/\omega^2$ Note that the signals obtained in this work and shown in Figs. 1a,b) are much cleaner and cover a much wider spectral range than what obtained in [21] after carefully running a number of runs larger by about three orders of magnitudes, each with a periodic forcing of different frequency.

## 5. Conclusions

In this paper we have proposed to study the impact of stochastic perturbations in the form of additive or multiplicative noise to deterministic dynamical systems using the Ruelle response theory. In this regard, this contributions is quite different from other papers recently appeared in the literature aiming at describing the impact of perturbations to the deterministic components of chaotic systems undergoing also stochastic forcing.

We have shown that performing experiments on a given system with and without adding stochastic perturbation provides a new way to access information on its response to more general forcings. In the case of a single white noise component modulating a perturbative vector flow, in agreement with the fact that SRB measure feature stochastic stability, the impact of stochastic forcings on the expectation value of a general observable vanishes with vanishing intensity of the noise, and is proportional to the variance of the noise. Moreover, the difference between the expectation value of the power spectrum of an observable in the stochastically perturbed case and of the same observable in the unperturbed case is equal to the noise variance times the square of the modulus of the susceptibility describing the frequency-dependent response of the system to perturbations with the same pattern as the considered stochastic forcing. Using Kramers-Kronig theory, it is then possible to derive the real and imaginary part of the susceptibility and, via inverse Fourier Transform, its Green function, which allows us to project perturbations into the future with a finite and infinite time horizon. At practical level, with only one ensemble of runs for the perturbed and unperturbed model we can derive the susceptibility and the Green function for any desired observable. Therefore, the results exposed in this paper allow for bridging the response of the system to very fast perturbations (in the form of white noise) to its behaviour and predictability at all time scales. We have clarified some of these results by resorting to an example, namely by considering the "intensive energy" and "intensive momentum" observables for the Lorenz 96 model in standard and stochastically perturbed conditions, where a simple additive white noise is taken into account. We have found that the quality in terms of signal to noise of the obtained squared modulus of the susceptibility and its spectral range are much wider than what derived using directly the response theory with periodic perturbations, even if the computational cost is in the present case much lower. This suggests a very efficient general way to derive the susceptibility of a system bypassing the correct but somewhat cumbersome procedure shown in [21]. The direct application of the fluctuation-dissipation theorem to the unperturbed deterministic system is not possible for general deterministic non equilibrium steady state systems described by Axiom A flows. Therefore, the analysis of the internal fluctuations of the unperturbed system does not allow for obtaining the properties of the response to external perturbations. In physical terms, this marks the difference between quasi-equilibrium and non-equilibrium systems [21]. Instead, adding stochastic perturbations smoothens the invariant measure and thus allows for the applicability of the fluctuation-dissipation theorem [32]. This is the fundamental reasons why we are able to obtain the results presented in this paper.

Furthermore, we have generalised our results for rather general patterns of random forcing, from the case of various perturbative vectors flows each modulated by a white noise process, to noise terms with memory, up to the case of a space-time random field. Explicit formulas are provided for each relevant case analysed. As a general result, he power spectrum of the stochastically perturbed system is larger at all frequencies than the power spectrum of the unperturbed system, which seems explicable as resulting from the input of energy from an outside environment and its redistribution inside the system at all temporal scales. Furthermore, the difference between the power spectrum of the perturbed and that of the unperturbed system can be written as a sum of products of first order susceptibilities each describing the response of the system to one of the applied perturbative vector flow. The case of a stochastic forcing given by a general spatial-temporal field $X(x,t)$ is treated by resorting to a Schauder decomposition.

When considering the spectral properties of the response to stochastic forcings, the findings presented in this paper are strikingly similar to what would be derived in the much elementary case of systems possessing simple attractors such as fixed points. This is the case because we are only exploiting the formal properties of the response formula, which, as discussed in [18], are, to a great extent, seamless with respect to the geometric properties of the attractor, which instead enter in the definition of the invariant measure.

We hope that this paper may provide stimulation, on one side, for providing a more rigorous analysis of stochastic perturbations on complex systems like the climate's, and, on the other side, to further investigate the relevance of the Ruelle response theory and of its spectral counterpart based upon Kramers-Kronig relations for studying the response of general systems to perturbations. We are currently investigating how the results contained in this paper may be used for devising parameterisation schemes for multi-level systems and whether for it is possible to draw a robust link with the recently proposed approach of random attractors [33].

**Acknowledgments**

The author wishes to thank T. Kuna, J. Wouters, D. Faranda. The financial support of the ERC-FR7 project NAMASTE "Thermodynamics of the Climate System" is gratefully acknowledged.

a)

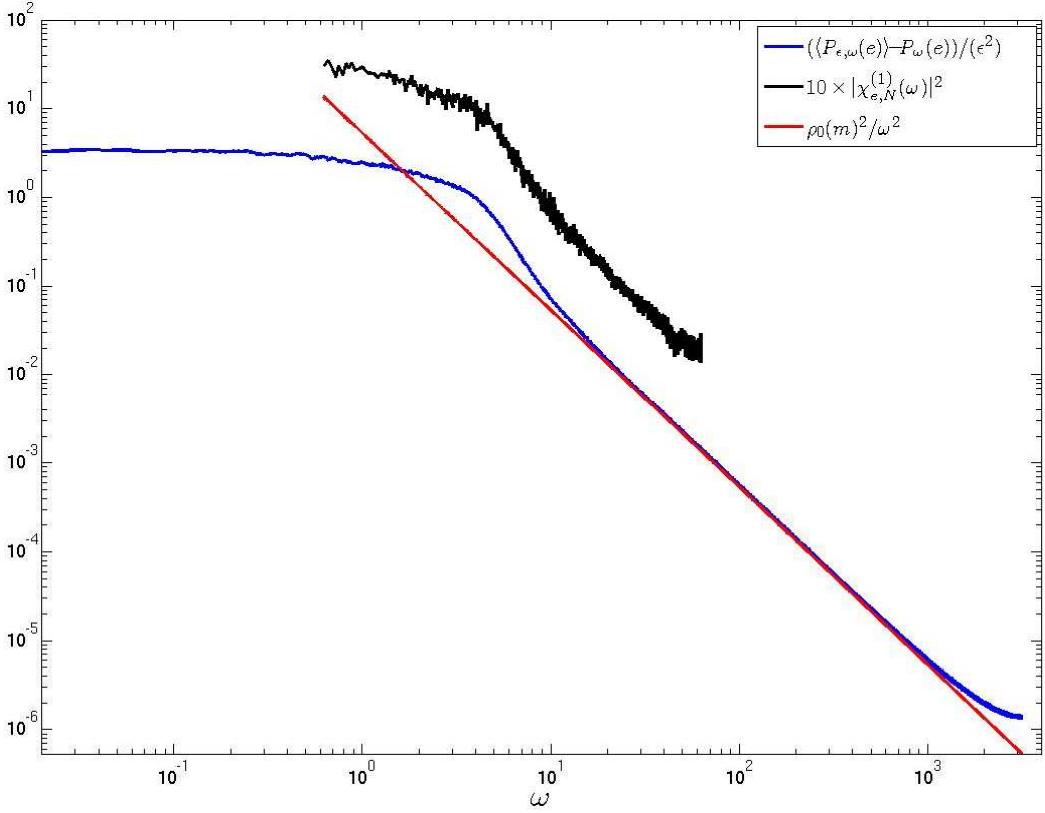

b)

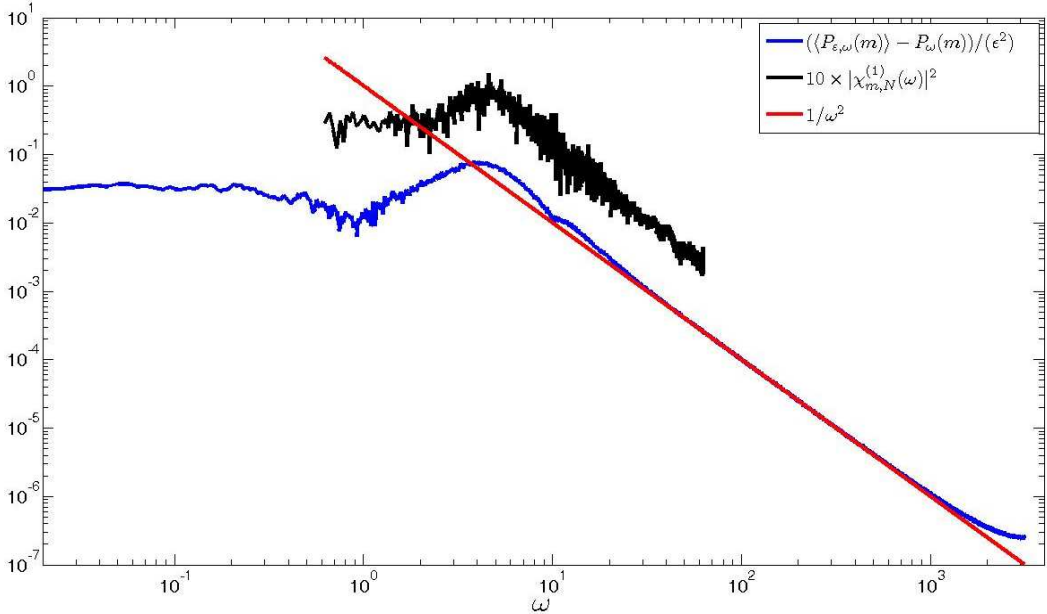

**Figure 1: How to reconstruct the square modulus of the susceptibility using stochastic perturbations. a) Blue line: evaluation of $\left|\chi_{e,N}^{(1)}(\omega)\right|^2$ via analysis of the impact of stochastic forcing using the formula given in Eq. (8). Black line: direct evaluation of $\left|\chi_{e,N}^{(1)}(\omega)\right|^2$ (data taken from [21]). The curve has been shifted (see legend) to improve readability. Red line: asymptotic behaviour derived analytically in [21]. b) Same as a), but for the function $\left|\chi_{m,N}^{(1)}(\omega)\right|^2$.**